\documentclass[prd,aps,showpacs,nofootinbib,preprint,eqsecnum]{revtex4}
\usepackage{graphicx,color,amsmath,amsxtra}
\usepackage{epsf}
\usepackage{amssymb}
\usepackage{enumerate}
\usepackage{hhline}
\usepackage{array}
\usepackage{tabularx}

\newcommand{\be}{\begin{equation}}
\newcommand{\ee}{\end{equation}}
\newcommand{\bea}{\begin{eqnarray}}
\newcommand{\eea}{\end{eqnarray}}
\newcommand{\beaa}{\begin{eqnarray*}}
\newcommand{\eeaa}{\end{eqnarray*}}

\newcommand{\e}{\mathrm{e}}

\newcommand{\Eqn}[1]{&\hspace{-0.2em}#1\hspace{-0.2em}&}
\newcommand{\abs}[1]{\vert{#1}\vert}
\def\Vec#1{\mbox{\boldmath $#1$}}
\def\Vecs#1{\mbox{\boldmath\tiny $#1$}}
\def\Lap{{\mathop{\Delta}\limits^{(3)}}}
\def\be{\begin{equation}}
\def\ee{\end{equation}}
\def\bea{\begin{eqnarray}}
\def\eea{\end{eqnarray}}

\def\e{\mathrm{e}}
\begin{document}

\title{Generation of large-scale magnetic fields from inflation 
in teleparallelism
}

\author{Kazuharu Bamba$^{1, 3}$\footnote{E-mail address: 
bamba@kmi.nagoya-u.ac.jp}, 
Chao-Qiang Geng$^{2,3,4}$\footnote{E-mail address: geng@phys.nthu.edu.tw} 
and 
Ling-Wei Luo$^{2,4}$\footnote{E-mail address: d9622508@oz.nthu.edu.tw} 
}
\affiliation{
$^1$Kobayashi-Maskawa Institute for the Origin of Particles and the
Universe,
Nagoya University, Nagoya 464-8602, Japan
\\
$^2$Department of Physics, National Tsing Hua University, Hsinchu, Taiwan 300 
\\ 
$^3$Physics Division, National Center for Theoretical Sciences, Hsinchu, Taiwan 300
\\
$^4$College of Mathematics \& Physics, Chongqing University of Posts \& Telecommunications, Chongqing, 400065, China
}

%\date{\today}

%%%%%%%%%%%%%%%%%%%%%
%  Abstract
%%%%%%%%%%%%%%%%%%%%%
\begin{abstract}
We explore the generation of large-scale magnetic fields from inflation in  teleparallelism, 
in which the gravitational theory is described by 
the torsion scalar instead of the scalar curvature in general relativity. 
In particular, we examine the case that the conformal invariance of 
the electromagnetic field during inflation is broken by 
a non-minimal gravitational coupling between 
the torsion scalar and the electromagnetic field. 
It is shown that for a power-law type coupling, 
the magnetic field on 1~Mpc scale with its 
strength of $\sim 10^{-9}$~G at the present time can be generated. 
\end{abstract}
%%%%%%%%%%%%%%%%%%%%%

%----------------------------
%\pacs{
%Keywords:
%}
%\preprint{}
%\hspace{13.0cm}
%----------------------------

\maketitle
%==============================================================================

%%%%%%%%%%%%%%%%%%%%%%%%%%%
%%%  Sec. I
%%%%%%%%%%%%%%%%%%%%%%%%%%%
\section{Introduction}

A number of recent cosmological observations, e.g., 
Type Ia Supernovae~\cite{SN1}, 
baryon acoustic oscillations~\cite{LSS}, 
large scale structure (LSS)~\cite{Eisenstein:2005su}, 
cosmic microwave background (CMB) radiation~\cite{WMAP, Komatsu:2008hk, Komatsu:2010fb}, 
and weak lensing~\cite{Jain:2003tba}, 
suggest the accelerated expansion of the current universe. 
%%%%%%%
To explain  the late time cosmic acceleration, 
there exist two main approaches: One is the 
introduction of the so-called ``dark energy'' 
(for reviews on dark energy, see, e.g.,~\cite{Copeland:2006wr, 
Caldwell:2009ix, Book-Amendola-Tsujikawa, Tsujikawa:2010sc, Li:2011sd, Kunz:2012aw, Bamba:2012cp}) 
and the other is the modification of gravity 
such as $f(R)$ gravity~\cite{Review-Nojiri-Odintsov, Sotiriou:2008rp, 
DeFelice:2010aj, Book-Capozziello-Faraoni, Tsujikawa:2010zza, Clifton:2011jh, Capozziello:2011et, Harko:2012ar, Capozziello:2012hm}. 

Recently, ``teleparallelism''~\cite{Teleparallelism} has attracted much 
attention 
as it can be considered as an alternative gravity theory to general relativity. 
Teleparallelism is formulated with the Weitzenb\"{o}ck connection, 
so that its action consists of the torsion scalar $T$ instead of 
the scalar curvature $R$  in general relativity
%,  constructed by 
with the Levi-Civita connection. 
%%%%%
It has been shown that  by introducing a scalar with the non-minimal 
coupling to  gravity in teleparallelism~\cite{TDE}, the late time cosmic acceleration
can be achieved.
Moreover, similar to $f(R)$ gravity, 
 the non-linear generalization of  the torsion scalar $T$, $i.e.$, 
%This is because it has been known that 
$f(T)$ gravity, 
%in which the action is described by an arbitrary function 
%$f(T)$ of the torsion scalar $T$ as is in $f(R)$ gravity, 
can account for inflation~\cite{Inflation-F-F} in the early universe 
as well as the cosmic acceleration 
in the late time~\cite{Bengochea:2008gz, L-BGL-f(T)}. 
Various aspects on $f(T)$ gravity have  been widely investigated 
in the literature (see, e.g.,~\cite{Bamba:2012cp} and the references therein). 
%%%%%

%%%%%%%%
On the other hand, 
according to astrophysical observations, 
it is well known that there exist magnetic fields 
with the strength  $\sim 10^{-6}$~G and
the coherence scale  $1$--$10$~kpc. 
Also in clusters of galaxies, 
large-scale magnetic fields are observed,
whose strengths are
$10^{-7}$ -- $10^{-6}$~G and the coherence scales are 
estimated as $10$~kpc--$1$~Mpc. However,
the origins of these cosmic magnetic fields, in particular 
the large-scale magnetic fields in clusters of galaxies 
have not been well understood yet
(for reviews on cosmic magnetic fields, see, e.g.,~\cite{M-F-Reviews}). 
There are several generation mechanisms of the cosmic magnetic fields, 
such as those from astrophysical processes based on 
the plasma instability~\cite{Biermann1, PI}, 
cosmological phase transitions~\cite{PT}, and 
matter density perturbations 
before or at the recombination epoch~\cite{DP}. 
Indeed, 
it is not so easy for 
these mechanisms to generate 
the large-scale magnetic fields observed in clusters of galaxies 
only with the adiabatic compression and 
without any secondary amplification mechanism as 
the galactic dynamo~\cite{EParker}. 
%%%%%%%%
Thus, the most natural mechanism to produce 
the large-scale magnetic fields is considered to be 
electromagnetic quantum fluctuations during inflation~\cite{Turner:1987bw}, 
because the scale of the electromagnetic quantum fluctuations 
can be extended to that larger than the Hubble horizon by inflation. 

In Quantum Electrodynamics (QED) in the curved space-time, 
there can appear a non-minimal coupling 
of the scalar curvature to the electromagnetic field 
owing to one-loop vacuum-polarization effects~\cite{Drummond:1979pp}, 
so that the conformal invariance of the electromagnetic field can be 
broken by this coupling. 
This can yield the quantum fluctuation of the electromagnetic 
field during inflation, 
resulting in the large-scale magnetic field 
at the present universe~\cite{Turner:1987bw, N-G-C}. 
Such a breaking mechanism of the conformal invariance of the electromagnetic field is 
necessary to generate the quantum fluctuation of the electromagnetic 
field, because the Maxwell theory is 
conformally invariant and 
the Friedmann-Lema\^{i}tre-Robertson-Walker (FLRW) space-time 
is conformally flat~\cite{Parker:1968mv}\footnote{
It should be noted that 
this is true only for the flat FLRW space-time, but 
not for the FLRW background with spatial 
curvature, e.g., an open FLRW universe~\cite{NZC}. 
In addition, 
the breaking of the conformal flatness during inflation 
has also been studied in Ref.~\cite{Maroto:2000zu}. 
Furthermore, there exist arguments in terms of 
the back reaction effect of the magnetic field generated during 
inflation~\cite{Backreaction, Fujita:2012rb, Suyama:2012wh}.}.  
Consequently, a lot of breaking mechanisms of 
the conformal invariance of the electromagnetic field 
have been explored
(for a list of these breaking mechanisms, see, e.g., 
reviews in~\cite{M-F-Reviews} 
and references in~\cite{Ratra:1991bn, Lemoine:1995vj, Gasperini:1995dh, 
B-Y, Martin:2007ue, M-BS-B, Bamba:2008my, Bamba:2008hr, M-BGHK-HKBG}). 

In this paper, 
motivated by both astrophysical and cosmological observations, 
we study the generation of large-scale magnetic fields from inflation in teleparallelism. 
In particular, 
we introduce a non-minimal gravitational coupling of 
the torsion scalar $T$ to the electromagnetic field 
by analogy with such an interaction between gravity and electromagnetism 
in general relativity. 
As an illustration, we demonstrate that for the form of the coupling to be a power-law type, 
the magnetic field with its current 
strength of $\sim 10^{-9}$G on 1Mpc scale can be generated. 

%%%%%%%%
It should be remarked that for example, 
in Ref.~\cite{Ratra:1991bn} Ratra has investigated the case that 
the gauge kinetic term is coupled to the inflaton field.
%%%
In this work, however,
the observation that there can be 
a spectator field evolving during inflation 
is used essentially. 
This possibility has been scrutinized in a number of different works, e.g., 
Refs.~\cite{Giovannini:2001nh, Giovannini:2007rh, Giovannini:2009xa} 
by Giovannini. 
In particular, 
in Ref.~\cite{Giovannini:2001nh} 
%the possibility of 
a scale-invariant spectrum 
during the conventional inflation has been demonstrated in a specific model 
where the gauge coupling is not a function of the inflaton 
(in the latter case the flatness of the potential might be spoiled).  
%%%%%%%%
%%%%% Units %%%%%
We use units of $k_\mathrm{B} = c = \hbar = 1$ and denote the
gravitational constant $8 \pi G$ by 
${\kappa}^2 \equiv 8\pi/{M_{\mathrm{Pl}}}^2$ 
with the Planck mass of $M_{\mathrm{Pl}} = G^{-1/2} = 1.2 \times 10^{19}$GeV. 
%%%%%%%%%%%%%%%%%

The paper is organized as follows. 
In Sec.\ II, we explain the fundamental formulations in teleparallelism. 
In Sec.\ III, in a non-minimal $I(T)$-Maxwell theory, where $I(T)$ is 
an arbitrary function of the torsion scalar $T$, 
we investigate the generation of large-scale magnetic fields in inflationary cosmology. 
In Sec.\ IV, for a concrete model of a power-law type coupling between the torsion scalar 
and the Maxwell field, we estimate the current strength of 
the large-scale magnetic field. 
Finally, conclusions are presented in Sec.\ V.

%%%%%%%%%%%%%%%%%%%%%%%%%%%
%%%  Sec. II
%%%%%%%%%%%%%%%%%%%%%%%%%%%
\section{Teleparallelism}

We adopt 
orthonormal tetrad components $e_A (x^{\mu})$ in  teleparallelism, 
where an index $A$ runs over $0, 1, 2, 3$ for the 
tangent space at each point $x^{\mu}$ of the manifold. 
The relations between the metric $g^{\mu\nu}$ and orthonormal tetrad components 
are given by 
$
g_{\mu\nu}=\eta_{A B} e^A_\mu e^B_\nu 
$, 
where $\mu$ and $\nu$ are coordinate indices on the manifold and 
run over $0, 1, 2, 3$. 
Hence, $e_A^\mu$ form the tangent vector of the manifold. 
We define 
the torsion and contorsion tensors as $T^\rho_{\verb| |\mu\nu} \equiv e^\rho_A 
\left( \partial_\mu e^A_\nu - \partial_\nu e^A_\mu \right)$ 
and 
$K^{\mu\nu}_{\verb|  |\rho} 
\equiv -\left(1/2\right) 
\left(T^{\mu\nu}_{\verb|  |\rho} - T^{\nu \mu}_{\verb|  |\rho} - 
T_\rho^{\verb| |\mu\nu}\right)$, respectively. 
Using these tensors, we construct the torsion scalar 
$T \equiv S_\rho^{\verb| |\mu\nu} T^\rho_{\verb| |\mu\nu}$ 
with 
$S_\rho^{\verb| |\mu\nu} \equiv \left(1/2\right)
\left(K^{\mu\nu}_{\verb|  |\rho}+\delta^\mu_\rho \ 
T^{\alpha \nu}_{\verb|  |\alpha}-\delta^\nu_\rho \ 
T^{\alpha \mu}_{\verb|  |\alpha}\right)$. 
In general relativity, the Einstein-Hilbert action 
consists of the scalar curvature $R$. 
However, in  teleparallelism  
the torsion scalar 
$T$ is used to represent the teleparallel Lagrangian density. 
As a result, 
the action in  teleparallelism is described by  
\begin{equation}
S_\mathrm{Tel} = 
\int d^4x \abs{e} \left[ T/\left(2{\kappa}^2\right) 
+{\mathcal{L}}_{\mathrm{M}} \right], 
\label{eq:2.1}
\end{equation}
where $\abs{e}= \det \left(e^A_\mu \right)=\sqrt{-g}$ 
and ${\mathcal{L}}_{\mathrm{M}}$ is the Lagrangian of matter.  
%%%%%
The variation of the action $S_\mathrm{Tel}$ with respect to 
the vierbein vector fields $e_A^\mu$ leads to 
the gravitational field equation~\cite{Bengochea:2008gz}, given by 
%
%\begin{equation}
%
$
\left(1/e\right) \partial_\mu \left( eS_A^{\verb| |\mu\nu} \right) 
-e_A^\lambda T^\rho_{\verb| |\mu \lambda} S_\rho^{\verb| |\nu\mu} 
+\left(1/4\right) e_A^\nu T = \left({\kappa}^2/2\right) e_A^\rho 
{T^{(\mathrm{M})}}_\rho^{\verb| |\nu}
$,
%
%\label{eq:2.2}
%\end{equation}
%
where ${T^{(\mathrm{M})}}_\rho^{\verb| |\nu}$ 
is the energy-momentum tensor of matter.  
%%%%%

%%%%%
We take the flat 
Friedmann-Lema\^{i}tre-Robertson-Walker (FLRW) 
universe, whose metric is given by 
%
%\begin{equation}
$
ds^2 = dt^2 - a^2(t) d{\Vec{x}}^2 
= a^2 (\eta) \left( -d \eta^2 + d{\Vec{x}}^2 \right) 
$
%\label{eq:2.1}
%\end{equation}
%
with $a$ the scale factor and $\eta$ the conformal time. 
In this space-time, 
$g_{\mu \nu}= \mathrm{diag} (1, -a^2, -a^2, -a^2)$ and 
the tetrad components become $e^A_\mu = (1,a,a,a)$. 
With these relations, 
we find that the exact value of 
the torsion scalar is described by $T=-6H^2$, where
$H \equiv \dot{a}/a$ is the Hubble parameter 
with the dot being the time derivative of $\partial/\partial t$.

%%%%%%%%%%%%%%%%%%%
%%%  Sec. III
%%%%%%%%%%%%%%%%%%%
\section{Non-minimal $I(T)$-Maxwell theory}

In this section, we consider a non-minimal $I(T)$-Maxwell theory 
and examine the generation of large-scale magnetic fields in inflationary cosmology. 

%%%%%%%%%%%%%%%%%%%
%%%  Sec. III A
%%%%%%%%%%%%%%%%%%%
\subsection{Model of the electromagnetic sector}

The action describing a non-minimal $I(T)$-Maxwell theory 
is given by 
\begin{equation}
S = 
\int d^{4}x \abs{e} 
\left(-\frac{1}{4}I(T)\, F_{\mu\nu}F^{\mu\nu}\right)\,,
\label{eq:3.1}
\end{equation}
where 
%$\abs{e}= \det \left(e^A_\mu \right)=\sqrt{-g}$, 
$I(T)$ is an arbitrary function of the torsion scalar $T$ and
$F_{\mu\nu} = {\partial}_{\mu}A_{\nu} - {\partial}_{\nu}A_{\mu}$ 
with $A_{\mu}$ the $U(1)$ gauge field is 
the electromagnetic field-strength tensor. 
It follows from the action in Eq.~(\ref{eq:3.1}) that 
the electromagnetic field equation is derived as 
\begin{equation}
-\frac{1}{\sqrt{-g}}{\partial}_{\mu} 
\left[ \sqrt{-g} I(T) F^{\mu\nu} 
\right] = 0\,. 
\label{eq:3.2}  
\end{equation}
In the FLRW background
with the Coulomb gauge 
$A_0(t,\Vec{x}) = 0$ and ${\partial}^jA_j(t,\Vec{x}) =0$, 
the equation of motion for the $U(1)$ gauge field 
is written as
\begin{equation}
\ddot{A_i}(t,\Vec{x}) 
+ \left( H + \frac{\dot{I}(T)}{I(T)} 
\right) \dot{A_i}(t,\Vec{x}) 
- \frac{1}{a^2}\Lap\, A_i(t,\Vec{x}) = 0\,, 
\label{eq:3.3}
\end{equation}
where $\Lap$ is the Laplacian in three-dimensional space. 

%%%%%%%%
We remark that 
%we have assumed that 
although 
%the coupling between
the electromagnetic field coupled (preferentially) to  the axial vector 
part of the torsion tensor 
%Indeed, such a coupling
 is more natural,
 we do not consider it in this study
%  However, 
since the resulting large-scale magnetic field is too small.
%this coupling would lead to a much smaller large-scale magnetic field, 
%we do not take into account the coupling explicitly. 
%%%%%%%
%%%%%%%
In addition, 
%related to the point above, 
%it should be 
we emphasize that 
the strength of the generated magnetic field is 
more related to the nature of the coupling $I$ to the gauge kinetic term 
than to the torsion scalar itself. 
%%%%%%%%

%%%%%%%%
It is important to explicitly state that the key element to generate 
the large-scale magnetic fields in the quasi-de Sitter phase of expansion, 
i.e., inflation in the early universe, 
is the scalar coupling between $I$ and the kinetic term of the electromagnetic 
field, as seen in Eq.~(\ref{eq:3.1}). 
In general, the coupling term $I(\eta)$ may be a function of various scalar 
degrees of freedom existing in a model, e.g., the inflaton or 
the dilaton field or a dynamic gauge coupling. 
As a result, $I(\eta)$ can be a function of a spectator field evolving during 
the inflationary epoch. 
In this case, there is no connection between the evolution of $I$ and the gauge coupling. Therefore, the physical features of the various models are 
different. 
In other words, in principle $I$ can be an arbitrary function of 
some non-trivial background fields. 
On the other hand, in the case of bouncing models 
some of these ideas are preferentially realized, 
whereas other models are consistent with the standard inflationary paradigm. 
%As a consequence, i
In this work, instead of concentrating on a specific mechanism for 
inflation, we execute a model-independent analysis 
on the generation of large-scale magnetic fields 
through the breaking of the conformal invariance of the 
electromagnetic field due to the coupling of $I$ in
in $-\left(1/4 \right) I F_{\mu\nu}F^{\mu\nu}$ in Eq.~(\ref{eq:3.1}). 
%%%%%%%%

%%%%%%%%
We also describe the realization of inflation. 
In this work, we suppose that the generic slow-roll inflation is realized 
without identifying the specific mechanism to lead to inflation. 
There are various possibilities to realize inflation. 
For example, one can introduce the inflaton coupling to 
the electromagnetic field~\cite{Ratra:1991bn, Martin:2007ue}, 
which can be considered to be the dilaton~\cite{Gasperini:1995dh}, 
or both inflaton and  dilaton fields~\cite{B-Y}.
%, only which couples to the electromagnetic field~\cite{B-Y}. 
Namely, if the dilaton field is not responsible for inflation, 
the coupling $I$ can be a function of the inflaton field. 
Furthermore, in these cases forms of the inflaton potential are assumed to be 
flat enough to realize the slow-roll inflation, namely, the quasi exponential 
expansion of the universe. 
Concrete demonstrations have been investigated 
in Refs.~\cite{Giovannini:2001nh, Giovannini:2007rh, Giovannini:2009xa}. 
%%%%%%%%

%%%%%%%%%%%%%%%%%%%
%%%  Sec. III B
%%%%%%%%%%%%%%%%%%%
\subsection{Quantization}

%Next, 
We now execute the quantization of $A_\mu (t,\Vec{x})$. 
{}From the action of the electromagnetic fields in Eq.~(\ref{eq:3.1}), 
we find that the canonical momenta conjugate to $A_{\mu}(t,\Vec{x})$ 
become 
%\begin{eqnarray}
${\pi}_0 = 0$, 
%\hspace{5mm} 
$
{\pi}_{i} = I a(t) \dot{A_i}(t,\Vec{x})
$. 
%\label{eq:3.1} 
%\end{eqnarray}
The canonical commutation relation 
between $A_i(t,\Vec{x})$ and ${\pi}_{j}(t,\Vec{x})$ 
is given by 
%\begin{eqnarray} 
$
\left[ \, A_i(t,\Vec{x}), {\pi}_{j}(t,\Vec{y}) \, \right] = i 
\int d^3 k (2\pi)^{-3} 
\e^{i \Vecs{k} \cdot \left( \Vecs{x} - \Vecs{y} \right)}
        \left( {\delta}_{ij} - k_i k_j/k^2 \right)
$ with $\Vec{k}$ being the comoving wave number and $k=|\Vec{k}|$. 
By imposing this relation, $A_i(t,\Vec{x})$ is described as 
%
%\begin{equation} 
$
A_i(t,\Vec{x}) = \int d^3 k (2\pi)^{-3/2} 
\sum_{\sigma=1,2}\left[ \, \hat{b}(\Vec{k},\sigma) 
        \epsilon_i(\Vec{k},\sigma)A(t,k) \e^{i \Vecs{k} \cdot \Vecs{x} }
       + {\hat{b}}^{\dagger}(\Vec{k},\sigma)
       \epsilon_i^*(\Vec{k},\sigma)
         {A^*}(t,k) \e^{-i \Vecs{k} \cdot \Vecs{x}} \, \right]
$,
where $\epsilon_i(\Vec{k},\sigma)$ ($\sigma=1,2$) 
stand for the two orthonormal transverse polarization 
vectors, and
%Moreover, 
%$\hat{b}(\Vec{k},\sigma)$ and 
${\hat{b}}^{(\dagger)}(\Vec{k},\sigma)$ 
is the annihilation (creation) operator, 
satisfying the relations 
$
\left[ \, \hat{b}(\Vec{k},\sigma),
 {\hat{b}}^{\dagger}({\Vec{k}}^{\prime},\sigma') \, \right] = 
\delta_{\sigma,\sigma'}
{\delta}^3 (\Vec{k}-{\Vec{k}}^{\prime})
$ 
and 
$
\left[ \, \hat{b}(\Vec{k},\sigma),
 \hat{b}({\Vec{k}}^{\prime},\sigma') \, \right] = 
\left[ \, {\hat{b}}^{\dagger}(\Vec{k},\sigma),
 {\hat{b}}^{\dagger}({\Vec{k}}^{\prime},\sigma') \, \right] = 0 
$. 
It follows from Eq.~(\ref{eq:3.3}) that 
the Fourier mode $A(k,t)$ obeys 
$
\ddot{A}(k,t) + \left( H + \dot{I}/I \right) \dot{A}(k,t) + 
\left(k^2/a^2\right) A(k,t) = 0
$ 
together with the normalization condition, 
$
A(k,t){\dot{A}}^{*}(k,t) - {\dot{A}}(k,t){A^{*}}(k,t)
= i/\left(I a\right)
$. 
If we use the conformal time $\eta$, this equation 
is rewritten as 
$A^{\prime \prime}(k,\eta) + 
\left( I^{\prime}/I \right) A^{\prime}(k,\eta) 
+ k^2 {A}(k,\eta) = 0
$, where 
the prime denotes the derivative in terms of $\eta$ 
as $\partial/\partial \eta$.

%%%%%%%%%%%%%%%%%%%
%%%  Sec. III C
%%%%%%%%%%%%%%%%%%%
\subsection{Procedure to obtain analytic solutions}

With the WKB approximation on subhorizon scales and the long 
wavelength approximation on superhorizon scales and matching these
solutions at the horizon crossing, 
it is possible to acquire an analytic solution for 
this equation~\cite{M-BS-B} approximately. 
%Here, we consider that the slow-roll inflation occurs, 
In this case of the exact de Sitter background, 
we find $a=1/(-H\eta)$ with 
$H$ being the Hubble parameter during the de Sitter expansion, 
and  $-k\eta=1$ at the horizon-crossing when $H=k/a$. 
For subhorizon (superhorizon) scales, we have 
$k|\eta|\gg1$ ($k|\eta|\ll1$). 
This is considered to be sufficiently well defined also for the
general slow-roll inflation, i.e., nearly exponential inflation. 

Provided that in the short-wavelength 
limit of $k/\left(aH\right) \gg 1$ 
the vacuum asymptotically approaches the Minkowski vacuum, 
the WKB subhorizon solution reads 
$
A_{\mathrm{in}} (k,\eta) = 
\left(1/\sqrt{2k}\right) I^{-1/2} e^{-ik\eta}
$. 

While, with the long-wavelength expansion in terms of $k^2$, 
we can have the solution on superhorizon scales 
$A_{\mathrm{out}} (k,\eta)$. 
By matching this solution with the above WKB subhorizon solution 
at the horizon crossing time $\eta = \eta_k \approx 1/k$, 
we obtain the lowest order approximate solution of 
$A_{\mathrm{out}} (k,\eta)$~\cite{M-BS-B} as 
\begin{eqnarray} 
A_{\mathrm{out}} (k,\eta) \Eqn{=} 
A_1(k) + A_2(k) \int_{\eta}^{{\eta}_{\mathrm{R}}} 
\frac{1}{I \left( \bar{\eta} \right)} d \bar{\eta}\,, 
\label{eq:3.4} \\ 
A_1(k) \Eqn{\equiv} 
\left. 
\frac{1}{\sqrt{2k}} I^{-1/2}  
\left[
1- \left( \frac{1}{2} I^{\prime} + i k I \right) 
\int_{\eta}^{{\eta}_{\mathrm{R}}} 
\frac{1}{I \left(\tilde{\bar{\eta}}\right)} 
d \tilde{\bar{\eta}} \right] \e^{-ik\eta} 
\right|_{\eta = \eta_k}\,, 
\label{eq:3.5} \\
A_2(k) \Eqn{\equiv} 
\left.
\frac{1}{\sqrt{2k}} I^{-1/2}  
\left( \frac{1}{2} I^{\prime} + i k I \right) 
\e^{-ik\eta} 
\right|_{\eta = \eta_k}\,.
\label{eq:3.6} 
\end{eqnarray}
We neglect the decaying mode solution, which is the second term of 
the right-hand side of  Eq.~(\ref{eq:3.4}). 
Equations (\ref{eq:3.5}) and (\ref{eq:3.6}) lead to 
$\left|A(k,\eta)\right|^2$ at the late times, given by 
\begin{equation}
\left|A(k,\eta)\right|^2 
= |A_1(k)|^2 
= \frac{1}{2kI(\eta_k)}
\left|1- \left(\frac{1}{2}\frac{I^{\prime}(\eta_k)}{kI(\eta_k)} 
+ i\right) k \int_{\eta_k}^{{\eta}_{\mathrm{R}}}
\frac{I(\eta_k)}{I \left(\tilde{\bar{\eta}} \right)}
d\tilde{\bar{\eta}}\,\right|^2\,. 
\label{eq:3.7}
\end{equation}
Here, we have supposed the instantaneous reheating after inflation 
and therefore, ${\eta}_{\mathrm{R}}$ is considered to be the conformal time 
at the reheating stage. 
By using the comoving magnetic field $B_i(t,\Vec{x})$, 
the proper magnetic field is expressed as 
$
{B_i}^{\mathrm{proper}}(t,\Vec{x})
    = a^{-1}B_i(t,\Vec{x}) = a^{-2}{\epsilon}_{ijk}{\partial}_j A_k(t,\Vec{x}),
$ 
where ${\epsilon}_{ijk}$ is the totally antisymmetric tensor 
with ${\epsilon}_{123}=1$. 
Accordingly, we find that 
the spectrum of the magnetic field is described as 
$
|{B}^{\mathrm{proper}}(k,\eta)|^2  
=2\left(k^2/a^4\right)|A(k,\eta)|^2
=2\left(k^2/a^4\right)|A_1(k)|^2
$, where we have taken into account the factor 2 originating from the two 
degrees of freedom for the polarization. 
In the Fourier space, the energy density of the magnetic field becomes 
${\rho}_B(k,\eta) = (1/2) |{B}^{\mathrm{proper}}(k,\eta)|^2 I(\eta)$. 
With multiplying this by the phase-space density $4 \pi k^3/(2\pi)^3$, 
we derive the energy density of the generated magnetic field per
unit logarithmic interval of $k$ as
\begin{equation}
\rho_B(k,\eta)\equiv 
\frac{1}{2} 
\frac{4\pi k^3}{(2\pi)^3}|{B}^{\mathrm{proper}}(k,\eta)|^2 I(\eta) 
=\frac{k|A_1(k)|^2}{2\pi^2}\frac{k^4}{a^4} I(\eta)\,.
\label{eq:3.8}
\end{equation}
Consequently, 
the density parameter of the magnetic field 
per unit logarithmic interval of $k$ 
and its spectral index 
are given by~\cite{M-BS-B} 
\begin{eqnarray}
\Omega_B(k,\eta) \Eqn{=} \frac{\rho_B(k,\eta_{\mathrm{R}})}
{\rho_\gamma(\eta_{\mathrm{R}})} 
\frac{I(\eta)}{I(\eta_{\mathrm{R}})} = 
\frac{k^4}{T_{\mathrm{R}}^4a_{\mathrm{R}}^4}\frac{15k|A_1(k)|^2}
{N_{\mathrm{eff}}\pi^4} I(\eta)\,,
\label{eq:3.9} \\ 
n_B \Eqn{\equiv} \frac{d\ln\Omega_B(k)}{d\ln k} =
4+\frac{d \ln k|A_1(k)|^2}{d \ln k}\,,
\label{eq:3.10} 
\end{eqnarray}
respectively, where, 
$
\rho_\gamma (\eta_{\mathrm{R}})=N_{\mathrm{eff}} \left(\pi^2 /30 \right) 
T_{\mathrm{R}}^4
$~\cite{Kolb and Turner} is the energy density of radiation at the reheating stage 
with the reheating temperature $T_{\mathrm{R}}$, 
$a_{\mathrm{R}}$ is the scale factor at $\eta = \eta_{\mathrm{R}}$ 
and $N_{\mathrm{eff}}$ is the effective massless degrees of freedom 
(e.g., for photons, $2$) 
thermalized at the reheating stage.

%%%%%%%%%%%%%%%%%%%%%%%%%%%
%%%  Sec. IV
%%%%%%%%%%%%%%%%%%%%%%%%%%%
\section{
Large-scale magnetic field generated in Teleparallelism
}

%%%%%%%%%%%%%%%%%%%%%%%%%%%
%%%  Sec. IV A
%%%%%%%%%%%%%%%%%%%%%%%%%%%
\subsection{
Current strength of the magnetic field 
%generated in the Teleparallelism
}

For the purpose of analyzing the strength of the magnetic field 
quantitatively, we examine the case of the specific form of $I(\eta)$, 
given by~\cite{M-BS-B}
\begin{equation}
I(\eta)=I_*\left(\frac{\eta}{\eta_*}\right)^{-\beta}\,,
\label{eq:4.1}
\end{equation}
where 
$\eta_*$ is some fiducial time at the inflationary stage, 
$I_*$ is the value of $I$ at $\eta=\eta_*$, 
and $\beta (>0)$ is a positive constant, whose positivity makes 
$I$ increase monotonically during inflation. 
For this form of $I$, 
$|A_1|^2$ in Eq.~(\ref{eq:3.7}) reads 
$
k|A_1|^2=\left(1/2I(\eta_k)\right) \left|1-\left(\beta+2i\right)/
\left[ 2 \left(\beta+1\right) \right] \right|^2 
\equiv \mathcal{A}/\left(2I(\eta_k)\right)
$, 
where $\mathcal{A} (= \mathcal{O}(1))$ is a constant  of the order of unity. 
By plugging this relation into Eqs.~(\ref{eq:3.9}) and (\ref{eq:3.10}), 
the density parameter of the magnetic field at the present time $\eta_0$ 
is expressed as 
$
\Omega_B(k,\eta_0)
= \left[ k^4/\left(T_{\mathrm{R}}^4 a_{\mathrm{R}}^4 \right) \right] 
\left[ 15\mathcal{A}/\left( 2N_{\mathrm{eff}}\pi^4I_* \right) \right] 
\left(\eta_k/\eta_*\right)^\beta
$ 
with $n_B=4-\beta$, 
where we have used $I(\eta_k)\propto k^\beta$ and $I(\eta_0) = 1$. 
If $I_*$ is very small and the spectrum is nearly scale-invariant, 
i.e., $\beta \sim 4$, the resultant amplitude of the 
large-scale magnetic field becomes large. 
{}From Eq.~(\ref{eq:3.9}), we find that 
the current density parameter of the magnetic field is 
described by~\cite{M-BS-B} 
\begin{equation}
\Omega_B(k,\eta_0)
= \mathcal{A} \frac{N_\mathrm{eff}}{1080}
\left(\frac{T_{\mathrm{R}}}{\tilde{M}_\mathrm{Pl}}\right)^4\left(-k\eta_{\mathrm{R}}\right)^{4-\beta}
\frac{1}{I(\eta_{\mathrm{R}})}\,.
\label{eq:4.2}
\end{equation}
Here, we have used 
the relation 
$a_{\mathrm{R}}^2\eta_{\mathrm{R}}^2 \approx H_{\mathrm{R}}^{-2}$ 
with $a_{\mathrm{R}}$ and $H_{\mathrm{R}}$ being 
the scale factor and the Hubble parameter at the reheating stage, 
respectively, 
and the Friedmann equation $3H_{\mathrm{R}}^2/ = 
\rho_\gamma (\eta_{\mathrm{R}}) /\tilde{M}_\mathrm{Pl}^2$ at the reheating stage, 
where $\tilde{M}_\mathrm{Pl} = M_\mathrm{Pl}/\sqrt{8\pi} = 1/\kappa$. 
We can further rewrite the term $\left(-k\eta_{\mathrm{R}}\right)$ 
as~\cite{Bamba:2008my} 
\begin{eqnarray} 
-k\eta_{\mathrm{R}} 
= \frac{k}{a_{\mathrm{R}}H_{\mathrm{R}}} \Eqn{\simeq} 
\left(\frac{1.88}{h}\right) 10^4 \left(\frac{L}{[\mathrm{Mpc}]}\right)
\left(\frac{T_{\mathrm{R}}}{T_0}\right)\left(\frac{H_0}{H_{\mathrm{R}}}\right) 
\label{eq:4.3} \\ 
\Eqn{=}
5.1 \times 10^{-25} N_\mathrm{eff}^{-1/2} 
\left(\frac{\tilde{M}_\mathrm{Pl}}{T_{\mathrm{R}}}\right) 
\left(\frac{L}{[\mathrm{Mpc}]}\right)^{-1}\,.
\label{eq:4.4} 
\end{eqnarray}
In deriving Eq.~(\ref{eq:4.3}), we have used 
$H_0^{-1}=3.0 \times 10^{3}\,h^{-1}$\,Mpc 
and $T \propto a^{-1}$, which leads to 
$\left(a_0/a_{\mathrm{R}}\right)=\left(T_{\mathrm{R}}/T_0\right)$. 
Moreover, in analyzing Eq.~(\ref{eq:4.4}), 
we have adopted the Friedmann equation
$3H_{\mathrm{R}}^2/ = 
\rho_\gamma (\eta_{\mathrm{R}}) /\tilde{M}_\mathrm{Pl}^2$ 
with 
$
\rho_\gamma (\eta_{\mathrm{R}})=N_{\mathrm{eff}} \left(\pi^2 /30 \right) 
T_{\mathrm{R}}^4
$, 
$T_0 = 2.73$\,K and $H_0=2.47 h \times 10^{-29}$\,K~\cite{Kolb and Turner} 
%$H_{0} = 2.1 h \times 10^{-42} \, \mathrm{GeV}$~\cite{Kolb-Turner}
with $h = 0.7$~\cite{Komatsu:2010fb, Freedman:2000cf, Riess:2009pu}. 
Since the current amplitude of the magnetic field is given by 
$|B (\eta_0)|^2 = 2 \rho_B (\eta_0) = 2 \Omega_B (\eta_0,k)\,\rho_\gamma 
(\eta_0)$, 
with $\rho_\gamma (\eta_0) \simeq 2 \times 10^{-51}\,\mathrm{GeV}^4$ 
and $1\,\mathrm{G}=1.95 \times 10^{-20}\,\mathrm{GeV}^2$ 
we find~\cite{Bamba:2008my}
%
%\begin{equation} 
%|B(\eta_0, L)| = 2.7 \left( 5.1 \right)^{-\beta/2} \times 10^{-56+12.5\beta} 
%N_\mathrm{eff}^{\left(\beta-2\right)/4} 
%\sqrt{ \mathcal{A} \frac{I(\eta_0)}{I(\eta_{\mathrm{R}})} } 
%\left( \frac{T_{\mathrm{R}}}{\tilde{M}_\mathrm{Pl}} \right)^{\beta/2}
%\left( \frac{L}{[\mathrm{Mpc}]} \right)^{\beta/2-2} \, \mathrm{G}\,.
%\label{eq:4.5}
%\end{equation}
%
%
\begin{equation} 
|B(\eta_0, L)| = 2.7 \left[  \frac{7.2}{(5.1)^4 \pi} \right]^{\beta/8} 
\times 10^{-56+51\beta/4} 
N_\mathrm{eff}^{\left(\beta-4\right)/8} 
\sqrt{ \mathcal{A} \frac{I(\eta_0)}{I(\eta_{\mathrm{R}})} } 
\left( \frac{H_{\mathrm{R}}}{M_\mathrm{Pl}} \right)^{\beta/4}
\left( \frac{L}{[\mathrm{Mpc}]} \right)^{\beta/2-2} \, \mathrm{G}\,.
\label{eq:4.5}
\end{equation}
We note that the reheating temperature $T_{\mathrm{R}}$ is 
described  by using the Hubble parameter at the end of inflation, 
namely, instantaneous reheating stage, 
$H_{\mathrm{R}}$ 
as $T_{\mathrm{R}} = \left[ 90/ \left(8 \pi^3 N_\mathrm{eff} \right) 
\right]^{1/4} \sqrt{M_\mathrm{Pl} H_{\mathrm{R}}}$. 
%
%Thus, we have 
%$\left( T_{\mathrm{R}}/ \tilde{M}_\mathrm{Pl} \right) = 
%\left[720/\left(\pi N_\mathrm{eff}\right)\right]^{1/4} 
%\sqrt{H_{\mathrm{R}}/M_\mathrm{Pl}}$. 
%
Furthermore, 
there exists the upper limit of $H_{\mathrm{R}}$ 
from tensor perturbations. 
With the Wilkinson Microwave Anisotropy Probe (WMAP) five year data 
in terms of the anisotropy of the CMB radiation~\cite{Komatsu:2008hk}, 
we have $H_\mathrm{R} < 6.0 \times 10^{14}$GeV~\cite{C-H}.

%%%%%%%%%%%%%%%%%%%%%%%%%%%
%%%  Sec. IV B
%%%%%%%%%%%%%%%%%%%%%%%%%%%
\subsection{
Estimation of the 
current strength of the large-scale magnetic field 
}

We suppose that power-law inflation occurs, in which the scale factor is 
given by 
\begin{equation}
a =a_0 \left( \frac{t}{t_0} \right)^p\,, 
\label{eq:4.6}
\end{equation}
with $p \gg 1$, where $a_0$ and 
$t_0$ are constants. The larger the value of $p$ is, the closer power-law inflation 
goes to exponential inflation. 
In this case, with the relation $\eta = \int \left( 1/a \right) dt$, we get
\begin{equation}
%\left( t/t_0 \right) 
\frac{t}{t_0} 
= \left[ a_0 t_0 \left( p-1 \right) \left( -\eta \right)
\right]^{-1/\left( p-1 \right)}\,.
\label{eq:4.7}
\end{equation}
%

%%%%%%%%%%%%%%%%%%%%%%%%%%%
%%%  Sec. IV B 1
%%%%%%%%%%%%%%%%%%%%%%%%%%%
%\subsubsection{Power-law type coupling}

%First, 
We examine the case of a power-law type coupling as 
\begin{equation} 
I(T) = \left( \frac{T}{T_0} \right)^{n}\,, 
\label{eq:4.8}
\end{equation}
where $T_0$ is a current value of $T$ and 
$n (\neq 0)$ is a non-zero constant. 
In this case, by using $T= -6H^2$, $H= p/t$ and Eq.~(\ref{eq:4.7}), 
we obtain 
\begin{equation*} 
I(\eta) = \left(-6/T_0 \right)^n \left( p/t_0 \right)^{2n} 
\left[ a_0 t_0 \left( p-1 \right) \right]^{2n/\left( p-1 \right)} 
\left( -\eta \right)^{2n/\left( p-1 \right)}.
\end{equation*}  
By comparing this equation with Eq.~(\ref{eq:4.1}), we acquire 
\begin{equation} 
\beta=-\frac{2 n}{p-1}\,. 
\label{eq:4.9}
\end{equation}
For $N_\mathrm{eff} = 100$, $H_{\mathrm{R}} = 1.0 \times 10^{14}$GeV 
($T_{\mathrm{R}} = 8.6 \times 10^{15}$GeV), 
$L = 1$Mpc, $\mathcal{A} = 1$, $I(\eta_{\mathrm{R}}) = I(\eta_0)$, 
and $\beta = 4.2$, which can be realized for $p=10$ and $n=-18.9$, 
we have
\begin{equation}
|B(\eta_0, L = 1\,\mathrm{Mpc})| = 2.5 \times 10^{-9} \, \mathrm{G}\,. 
\end{equation}
%
%Since $p \gg 1$, it follows from Eq.~(\ref{eq:4.9}) that 
%$\beta \simeq -2n/p$. For example, if $p=100$, for $\beta = 4.2
Similarly, for the above values except 
$H_{\mathrm{R}} = 1.0 \times 10^{10}$GeV 
($T_{\mathrm{R}} = 8.6 \times 10^{13}$GeV) 
and $\beta = 4.6$, met for $p=10$ and $n=-19.7$, 
we obtain
\begin{equation}
|B(\eta_0, L = 1\,\mathrm{Mpc})| = 2.3 \times 10^{-9} \, \mathrm{G}\,.
\end{equation}
Here, it should be mentioned that in order to demonstrate the 
estimation of the generated magnetic field strength at the present time, 
we have considered the case in which the non-minimal gravitational coupling 
of the electromagnetic field $I(T)$ changes in time only during 
inflation, whereas it does not evolve any more, i.e., 
$I(\eta_{\mathrm{R}}) = I(\eta_0)$, after the instantaneous reheating stage following inflation. 

%%%%%%%%
Finally, we compare our results with the analysis executed in Ref.~\cite{Giovannini:2009xa}. 
The parameter of $\beta $ in our study corresponds to that of $\nu$ in Ref.~\cite{Giovannini:2009xa}.
In particular, $\beta = 4.2$ in the present work correlate with $\nu = 2.6$.
%The case of $\beta = 4.2$ in the present study corresponds to the case $\nu = 2.6$, 
%which is a parameter of the model in Ref.~\cite{Giovannini:2009xa}, 
%where 
Note that the scale-invariant spectrum of the magnetic fields 
is obtained for $\nu = 5/2$ in Ref.~\cite{Giovannini:2009xa},
In comparison with the analysis in Ref.~\cite{Giovannini:2009xa}, 
 for $\nu = 2.6$ the resultant strength of the magnetic field is 
estimated as $5.4 \times 10^{-9}$G, 
whereas in the scale-invariant limit 
the magnetic field would be $1.4410 \times 10^{-11}$ G. 
These figures may change depending on the assumptions on the reheating stage. 
Hence, we suppose the sudden (i.e., spontaneous) reheating 
where all the energy density of the inflaton can safely be assumed to be released into the energy density of the radiation.  
In this sense, for the case where $H_\mathrm{R} = 10^{10}$ GeV 
($T_{\mathrm{R}} = 8.6 \times 10^{13}$GeV) with 
the above values such as $\beta = 4.2$ (i.e., $p=10$ and $n=-18.9$), 
we find 
\begin{equation}
|B(\eta_0, L = 1\,\mathrm{Mpc})| = 1.6 \times 10^{-13}~\mathrm{G}\,.
\end{equation}
Clearly, this strength can satisfy the scale-invariant limit, 
namely, less than $1.4410 \times 10^{-11}$~G. 
%%%%%%%%

%%%%%%%%%%%%%%%%%%%
%%%  Sec. V
%%%%%%%%%%%%%%%%%%%
\section{Conclusions}

We have investigated the generation of large-scale magnetic fields 
in inflationary cosmology in the context of teleparallelism. 
We have examined a non-minimal gravitational coupling of the torsion scalar 
to the electromagnetic field, which breaks its 
conformal invariance and hence, the quantum fluctuations of the 
electromagnetic field can be produced during inflation. 
It has explicitly illustrated that 
if the form of the coupling is a power-law type, 
the magnetic field with its
strength of $\sim 10^{-9}$G and the coherence scale of 1Mpc 
at the present time can be generated. 
This field strength is enough to account for the large-scale 
magnetic fields observed in clusters of galaxies only through 
the adiabatic compression during the construction of the large scale 
structure of the universe without the dynamo amplification mechanism. 

%%%%%%%%
Finally, 
we remark that 
the resultant field strength of $\sim 10^{-9}$G on 1Mpc scale 
is compatible with the upper limit of 
$\sim 2$--$6 \times 10^{-9}$G obtained 
from the observation of CMB radiation~\cite{CMB-Limit, Giovannini:2008df} 
as well as that of being smaller than 
$4.8 \times 10^{-9} \mathrm{G}$ 
from CMB radiation on the present strength 
with scales larger than the present horizon~\cite{Barrow:1997mj}. 
%%%%%
There also exist constraints on the strength of the large-scale magnetic 
fields 
from the matter density fluctuation parameter $\sigma_8$~\cite{YIKM}, 
the fifth science (S5) run of laser interferometer gravitational-wave 
observatory (LIGO)~\cite{Wang:2008vp}, 
Chandra X-ray galaxy cluster survey and Sunyaev-Zel'divich (S-Z) 
survey~\cite{Tashiro:2010st}, 
which are compatible with or weaker than those from CMB. 
Incidentally, 
generic features of the spectrum of the large-scale magnetic fields 
generated at the inflationary stage 
have been investigated in Ref.~\cite{Bamba:2007hm}. 
%%%%%
Moreover, it is also known that 
from the Big Bang Nucleosynthesis (BBN), 
there are limits on the primordial magnetic fields 
The constraint on the current strength of the magnetic fields on 
the BBN horizon scale $\sim 9.8 \times 10^{-5} h^{-1}\mathrm{Mpc}$, 
where $h=0.7$~\cite{Freedman:2000cf}, 
is smaller than $10^{-6}$G~\cite{BBN}. 
%%%%%
Furthermore, it is meaningful to note that 
the large-scale magnetic fields with 
the strength $\sim 4 \times 10^{-11}$--$10^{-10}$G 
at the present time can be observed~\cite{Test} by 
various future polarization experiments on CMB radiation, 
e.g., PLANCK~\cite{Planck-1, Planck-2}, 
QUIET~\cite{QUIET-1, Samtleben:2008rb}, 
B-Pol~\cite{B-Pol} and LiteBIRD~\cite{LiteBIRD}. 
If such large-scale magnetic fields in void regions 
and/or inter galactic medium 
are detected, the possibility that those origin is 
the quantum fluctuations of the electromagnetic field generated 
at the inflationary stage would become higher. 
Thus, physics in the early universe including inflation may be 
understood through the future detection of the large-scale magnetic fields. 
%%%%%

%%%%%%%%%%%%%%%%%%%%%%%%
%%%  Acknowledgments
%%%%%%%%%%%%%%%%%%%%%%%%
\section*{Acknowledgments}
%%%%%%%%
K.B. would like to sincerely thank the 
very kind and warm hospitality 
at National Center for Theoretical Sciences and 
National Tsing Hua University 
very much, where this work was initiated. 
%%%%%%%%
This work was partially supported by National Center of Theoretical
Science and  National Science Council (NSC-98-2112-M-007-008-MY3 and
NSC-101-2112-M-007-006-MY3) of R.O.C.

%%%%%%%%%%%%%%%%%%%
%%%  Appendix
%%%%%%%%%%%%%%%%%%%
%\appendix
%%%%%%%%%%%%%%%%%%%
%%%  A.
%%%%%%%%%%%%%%%%%%%
%\section{} 

%In this appendix, we  

%%%%%%%%%%%%%%%%%%%%%%%%%%%%%%%%%
%% thebibliography environment
%%%%%%%%%%%%%%%%%%%%%%%%%%%%%%%%%

\end{document}